# Multicomponent CASSCF Revisited: Large Active Spaces are Needed for Qualitatively Accurate Protonic Densities

O. Jonathan Fajen, Kurt R. Brorsen[*]

Department of Chemistry, University of Missouri, Columbia, Missouri 65203, USA

**Abstract**

Multicomponent methods seek to treat select nuclei, typically protons, fully quantum mechanically and equivalent to the electrons of a chemical system. In such methods, it is well known that due to the neglect of electron-proton correlation, a Hartree-Fock (HF) description of the electron-proton interaction catastrophically fails leading to qualitatively incorrect protonic properties. In single-component quantum chemistry, the qualitative failure of HF is normally indicative of the need for multireference methods such as complete active space self-consistent field (CASSCF). While a multicomponent CASSCF method was implemented nearly twenty years ago, it is only able to perform calculations with very small active spaces (~$10^5$ multicomponent configurations). Therefore, in order to extend the realm of applicability of the multicomponent CASSCF method, this study derives and implements a new two-step multicomponent CASSCF method that uses multicomponent heat-bath configuration interaction for the configuration interaction step, enabling calculations with very large active spaces (up to 16 electrons in 48 orbitals). We find that large electronic active spaces are needed to obtain qualitatively accurate protonic densities for the HCN and FHF$^-$ molecules. Additionally, the multicomponent CASSCF method implemented here should have further applications for double-well protonic potentials and systems that are inherently electronically multireference.

## 1. Introduction

Multicomponent *ab initio* methods[1-5] are an emerging field of quantum chemistry that treat select nuclei, typically protons, fully quantum mechanically in a manner identical to that of the electrons of a system. These methods in their modern formulation have existed for over twenty years, but it was not until recently that accurate protonic properties could be obtained for molecular systems. This was largely because of the difficulty in correctly describing the attractive electron-proton interaction. If the interaction is treated at the mean-field or Hartree-Fock (HF) level, the ground-state protonic orbital catastrophically over localizes, which leads to a qualitatively incorrect description of most protonic properties.

The failure of multicomponent HF[1-2] arises from its neglect of all electron-proton correlation and so the majority of multicomponent method development over the past two decades has focused on its inclusion. Unfortunately, due to the qualitatively incorrect nature of multicomponent Hartree-Fock, multicomponent extensions of standard single-component methodology such as truncated configuration interaction (CI) with single and doubles excitations[6] or Møller-Plesset second-order perturbation theory[7-9] (MP2) are not even qualitatively accurate as the multicomponent Hartree-Fock wave function is not a sufficiently good reference wave function. Multicomponent coupled cluster (CC) theory[6, 8-13] has been shown to give accurate protonic properties only if single excitations are included in the cluster operator, which is another indication of the poor quality of the multicomponent HF orbitals. Additionally, if excitations up to quadruples are included in a truncated CI expansion (CISDTQ), accurate protonic densities can be obtained.[14]

In the last few years, it has been realized that orbital-optimized multicomponent methods, where the electronic and protonic orbitals are optimized in the presence of electron-proton correlation, are able to compute accurate protonic properties, as they do not rely on the multicomponent HF orbitals except possibly as an initial guess. This has led not only to the first multicomponent method able to compute even qualitatively accurate protonic properties for realistically-sized chemical systems,[15-16] but also to the rapid development of additional multicomponent orbital-optimized methods using MP2,[17-18] CC theory,[18] and a variety of additional DFT functionals.[19-20] Using the results from these recent orbital-optimized studies, we have hypothesized that accurate protonic properties can be obtained using multicomponent generalizations of single-component quantum chemistry methods as long as the orbitals are optimized in the presence of electron-proton correlation.[17]

In single-component quantum chemistry, the failure of HF is normally indicative of the need for multireference methods such as complete active-space self-consistent field (CASSCF)[21] or density matrix renormalization group (DMRG).[22-23] A previous study[14] has shown that using a multicomponent HF wave function as a reference wave function introduces significant multireference character into the wave function, which would seem to indicate the need for multicomponent multireference methods. A multicomponent CASSCF method was derived and implemented nearly twenty years ago in the initial introduction of the nuclear-electronic orbital (NEO) framework,[2] which is one of the leading frameworks for multicomponent quantum chemistry. Neither the initial implementation nor follow up studies[24-25] were able to obtain qualitatively accurate protonic properties for realistic molecular systems. Additionally, we briefly note that a multicomponent DMRG method was recently derived[26] that in principle can describe multireference character accurately like multicomponent CASSCF.

Due to severe technical limitations, the only publicly available multicomponent CASSCF code, in GAMESS,[27] is unable to perform calculations with CI expansions greater than $\sim 10^5$ multicomponent configurations in our test calculations, which likely explains in part the small active spaces used in previous multicomponent CASSCF studies. To the best of our knowledge, the multicomponent CASSCF calculations with largest active spaces in the literature are 2 electrons in 16 orbitals with 1 proton in 20 orbitals for the HF molecule[2] and 4 electrons in 8 orbitals with 1 proton in 32 orbitals for the ClHCl⁻ molecule.[25] In the latter study, it was claimed that the improvement in the energetics for the multicomponent systems when going from an electronic active space of 4 electrons in 4 orbitals to 4 electrons in 6 orbitals was due to the larger active space including more dynamic electron-proton correlation. However, given the small size of both of these active spaces, it is unlikely that either of them included any significant amount of dynamic electron-proton correlation, which ultimately makes the truth of this claim uncertain at present.

Furthermore, recent multicomponent calculations have consistently demonstrated[5] the need for large electronic basis sets for the hydrogen atom such as cc-pv5z[28] and cc-pv6z[29] and large $8s8p8d$ or $8s8p8d8f$ even-tempered[30] protonic basis sets in order to obtain accurate protonic properties, though the recently introduced PB family of protonic basis sets[31] should reduce the size of the protonic basis sets in multicomponent calculations. Previous multicomponent CASSCF calculations used the 6-31G(*d,p*) electronic basis set[32] with a $2s2p2d$ or $2s6p$ nuclear basis set,[2, 25] which are likely too small to obtain accurate protonic properties even at the full CI level. Therefore, previous studies have yet to truly offer any insights into how large the active space of a multicomponent CASSCF calculation must be for accurate protonic properties.

In this study, in an effort to extend the applicability of the multicomponent CASSCF method and examine the size of the active space required to obtain accurate protonic properties, we derive and implement a new two-step multicomponent CASSCF method where the CI step uses the multicomponent heat-bath CI method[14, 33-34] to solve for the CI coefficients. The multicomponent heat-bath CI (HCI) method is a selected CI method previously used to perform multicomponent CISDTQ calculations.[14] In part, the present study can be considered a multicomponent generalization of the single-component heat-bath-CI SCF (HCISCF)[35] and adaptive-sampling-CI SCF[36-37] methods, the latter of which is another selected CI SCF method. We use a multicomponent generalization of the former to solve for the CI coefficients and a multicomponent generalization of the latter to optimize the orbital coefficients. We briefly note that a multicomponent adaptive-sampling-CI method has been previously derived in the context of calculating the many-body states of quantum nanostructures, but it did not include any orbital optimization.[38]

In single-component quantum chemistry, selected CI solvers have enabled the use of very large active spaces in CASSCF calculations such as 44 electrons in 44 orbitals for the HCISCF method[35] or 52 electrons in 52 orbitals in the adaptive-sampling-CI CASSCF method.[37] By using a multicomponent generalization of these methods, we are able to perform multicomponent CASSCF calculations using very large active spaces relative to the previous results in the literature. Additionally, our multicomponent CASSCF implementation differs from that of the previous multicomponent CASSCF implementation in that not all the protonic orbitals are included in the active space, which allows the assessment of how many active protonic orbitals are needed for accurate protonic properties and should increase the computational efficiency of the multicomponent CASSCF method.

As was discussed in a recent review article,[5] an accurate, computationally tractable multicomponent CASSCF method will also be widely applicable for future multicomponent studies. One of the most promising use cases for multicomponent methods is in describing proton-coupled electron transfer (PCET).[39-41] For such systems, the transferring proton must be treated quantum mechanically and commonly is in a double-well potential such that the ground-state protonic density is localized in multiple wells. It is widely assumed in the literature that the description of such a system will require multireference methods such as multicomponent CASSCF. Additionally, any PCET studies involving metals or other systems where the electronic structure is inherently multireference will require that the static electron-electron correlation be accurately included, which the multicomponent CASSCF method in this study is capable of doing in a manner identical to that of single-component CASSCF.

The paper is organized as follows. In Section 2, we introduce our implementation of the multicomponent CASSCF method and the multicomponent HCI method that is used as a CI solver. In Section 3, we benchmark the multicomponent CASSCF method on the HCN and FHF⁻ molecules and show that large active spaces are required to achieve a level of accuracy in the protonic density similar to that of previous multicomponent orbital-optimized methods. Finally, in Section 4, we conclude and briefly discuss the future utility of the multicomponent CASSCF method and possible future extensions.

## 2. Methodological Approach

In this Section, we first introduce the multicomponent HCI method that will be used as the CI solver in the two-step multicomponent CASSCF method before introducing the rest of the multicomponent CASSCF method including the equations for the orbital update step.

In the following, the letters $i, j$ denote doubly occupied or core electronic orbitals, $t, u, v,...$ denote active electronic orbitals, $k, l$ denote any occupied orbital, $a, b, c,...$ denote unoccupied or virtual electronic orbitals, and $p, q, r,...$ denote any type of orbitals. Protonic orbitals are denoted analogously, but with capital letters. Greek letters are used to index multicomponent configurations.

The multicomponent CASSCF method is derived for a single quantum proton as the likely application for the method is for a single transferring proton in a PCET system. Therefore, there are no core protonic orbitals in our implementation, and we treat the single proton as an alpha-spin proton. The generalization to multiple protons with a high-spin protonic wave function, which is typically assumed in the multiple proton case[42] in a multicomponent framework, is straightforward.

## 2.1 Multicomponent Heat-Bath Configuration Interaction

In multicomponent CI, the wave function, $|\Psi\rangle$, is written as a linear combination of multicomponent configurations, $|\Phi_\mu^{\text{elec}}\rangle|\Phi_\nu^{\text{prot}}\rangle$,

$$|\Psi\rangle = \sum_{\mu\nu} c_{\mu\nu} |\Phi_\mu^{\text{elec}}\rangle|\Phi_\nu^{\text{prot}}\rangle, \qquad (1)$$

where $|\Phi_\mu^{\text{elec}}\rangle$ or $|\Phi_\nu^{\text{prot}}\rangle$ is a Slater determinant of electronic or protonic orbitals, respectively. We will also write a multicomponent configuration as $|\Phi_\mu^{\text{elec}}\Phi_\nu^{\text{prot}}\rangle$. For a multicomponent system, the multicomponent Hamiltonian in atomic units is

$$\hat{H} = \sum_{pq} h_{pq}^{\text{elec}} \hat{E}_{pq} + \frac{1}{2}\sum_{pqrs}(pq|rs)\hat{E}_{pq,rs}$$
$$+ \sum_{PQ} h_{PQ}^{\text{prot}} \hat{E}_{PQ} - \frac{1}{2}\sum_{pqPQ}(pq|PQ)\hat{E}_{pq,PQ} + E_{\text{nuc}} \qquad (2)$$

where $\hat{E}_{pq}$ and $\hat{E}_{PQ}$ are the one-particle excitation operators:

$$\hat{E}_{pq} = \hat{E}_{pq}^{\alpha} + \hat{E}_{pq}^{\beta} = a_{p\alpha}^{\dagger} a_{q\alpha} + a_{p\beta}^{\dagger} a_{q\beta} \tag{3}$$

$$\hat{E}_{PQ} = \hat{E}_{PQ}^{\alpha} = a_{P\alpha}^{\dagger} a_{Q\alpha} \tag{4}$$

and $\hat{E}_{pq,rs}$ and $\hat{E}_{pq,PQ}$ are the two-particle excitation operators:

$$\hat{E}_{pq,rs} = \hat{E}_{pq} \hat{E}_{rs} - \delta_{qr} \hat{E}_{ps} \tag{5}$$

$$\hat{E}_{pq,PQ} = \hat{E}_{pq} \hat{E}_{PQ}, \tag{6}$$

and $E_{\text{nuc}}$ is the classical nuclear repulsion energy. The energy of the system is

$$\begin{aligned} E &= 2\sum_i h_{ii}^{\text{elec}} + \sum_{ij} \left[ 2(ii|jj) - (ij|ij) \right] + \sum_{tu} D_{tu} h_{tu}^{\text{elec}} + \frac{1}{2} \sum_{tuvw} D_{tu,vw} (tu|vw) \\ &+ \sum_{itu} D_{tu} \left[ (ii|tu) - \frac{1}{2}(it|iu) \right] + \sum_{TU} D_{TU} h_{TU}^{\text{prot}} - \sum_{tuTU} D_{tu,TU} (tu|TU) - 2\sum_{iTU} D_{TU} (ii|TU), \end{aligned} \tag{7}$$

where the 1-particle electronic and protonic reduced density matrices are defined, respectively, as

$$D_{pq} = \sum_{\mu\nu\lambda\sigma} c_{\mu\nu} c_{\lambda\sigma} \left\langle \Phi_\mu^{\text{elec}} \Phi_\nu^{\text{prot}} \middle| \hat{E}_{pq} \middle| \Phi_\lambda^{\text{elec}} \Phi_\sigma^{\text{prot}} \right\rangle, \tag{8}$$

$$D_{PQ} = \sum_{\mu\nu\lambda\sigma} c_{\mu\nu} c_{\lambda\sigma} \left\langle \Phi_\mu^{\text{elec}} \Phi_\nu^{\text{prot}} \middle| \hat{E}_{PQ} \middle| \Phi_\lambda^{\text{elec}} \Phi_\sigma^{\text{prot}} \right\rangle, \tag{9}$$

and the 2-particle electronic and electronic-protonic reduced density matrices are defined, respectively, as

$$D_{pq,rs} = \sum_{\mu\nu\lambda\sigma} c_{\mu\nu} c_{\lambda\sigma} \left\langle \Phi_\mu^{\text{elec}} \Phi_\nu^{\text{prot}} \middle| \hat{E}_{pq,rs} \middle| \Phi_\lambda^{\text{elec}} \Phi_\sigma^{\text{prot}} \right\rangle, \tag{10}$$

$$D_{pq,PQ} = \sum_{\mu\nu\lambda\sigma} c_{\mu\nu} c_{\lambda\sigma} \left\langle \Phi_\mu^{\text{elec}} \Phi_\nu^{\text{prot}} \middle| \hat{E}_{pq,PQ} \middle| \Phi_\lambda^{\text{elec}} \Phi_\sigma^{\text{prot}} \right\rangle. \tag{11}$$

The 2-particle electronic reduced density matrix is further symmetrized so that it exhibits the same eight-fold symmetry of the two-electron MO integrals.

Using the definitions of the wave function and Hamiltonian, multicomponent full CI and CASCI wave functions and energies can be obtain by constructing the Hamiltonian matrix using the multicomponent Slater-Condon rules[2] and then diagonalizing the Hamiltonian matrix.

However, the number of multicomponent configurations in Equation 1 increases rapidly with the size of the active space in a manner analogous to single-component CI. Additionally, when a full CI or CASCI calculation is performed, it is observed that most of the expansion coefficients in Equation 1 are close to zero indicating that they could be dropped from the expansion without any considerable loss of accuracy in the wave function. This concept has been explored under the name of "deadwood" in single-component full CI.[43]

The last few years have seen a renaissance in single-component quantum chemistry in full CI methods[44] and in particular of so-called selected CI methods[33-37, 45-51] that seek to keep the number of terms in the single-component analogue of Equation 1 small by the use of a ranking algorithm or selection criterion to include in the CI expansion only those electronic determinants that are likely to contribute significantly to the wave function. Recently, we have adapted one of these methods, HCI,[33-34] to a multicomponent formalism and used it to perform multicomponent CISDTQ calculations.[14]

Multicomponent HCI begins with the specification of an initial guess for the wave function in Equation 1. In this study, the guess will always be a single multicomponent configuration, but in principle, a small multicomponent CASCI guess could be used. This guess defines the current variational space. A multicomponent configuration, $\left|\Phi_\sigma^{elec}\Phi_\lambda^{nuc}\right\rangle$, is then added to the variational space if it satisfies the selection criterion:

$$\max_{\mu\nu}\left(\left|H_{\sigma\lambda,\mu\nu}c_{\mu\nu}\right|\right) > \varepsilon \tag{12}$$

where $H_{\sigma\lambda,\mu\nu}$ is the Hamiltonian matrix element between the multicomponent configurations, $\left|\Phi_\mu^{elec}\Phi_\nu^{prot}\right\rangle$ and $\left|\Phi_\sigma^{elec}\Phi_\lambda^{prot}\right\rangle$, $\varepsilon$ is an user-chosen cutoff value, and $\left|\Phi_\mu^{elec}\Phi_\nu^{prot}\right\rangle$ is a multicomponent configuration in the current variational space. After the new variational space has been found, the Hamiltonian is then constructed in the enlarged variational space basis and diagonalized to obtain a new set of expansion coefficients for the wave function. The process then repeats itself beginning with the enlargement of the variational space using Equation 12. Multicomponent HCI is terminated either when the number of multicomponent configurations added to the variational space is less than 0.01% of total number of multicomponent configurations in the current variational space or the energy difference between two successive iterations is less than $10^{-6}$ Ha. More details about the multicomponent heat-bath CI method such as the justification for the selection algorithm in Equation 12 can be found in a previous study.[14]

## 2.2 Multicomponent CASSCF

In multicomponent CASSCF, the wave function is still written as in multicomponent CI using Equation 1, but during the calculation the expansion coefficients in Equation 1 and the molecular orbital coefficients of the Slater determinants, $\left|\Phi_\mu^{elec}\right\rangle$ and $\left|\Phi_\nu^{prot}\right\rangle$, are optimized simultaneously. Similar to the previous multicomponent CASSCF implementation,[2] our multicomponent CASSCF implementation is a two-step method that decouples the CI step from the orbital-update step and uses a multicomponent generalization of the method of Chaban[52] for the orbital-update step. This updates the orbital coefficients using a quasi-Newton step with a diagonal orbital Hessian. A similar orbital update was also used in the single-component

CASSCF method using the adaptive-sampling CI method.[37] Our notation differs slightly from that of Chaban and the previous multicomponent CASSCF study as we follow the notation of Kreplin in their recent CASSCF implemenation.[53-54] In our two-step procedure, each macroiteration begins with solving the CI equation using the multicomponent HCI method to select a set of multicomponent configurations. The Hamiltonian is then constructed using this set and diagonalized to obtain an updated set of CI coefficients in Equation 1. After the final set of CI coefficients is obtained, a series of microiterations is performed with the CI coefficients held fixed during which the molecular orbital coefficients are optimized.

The change in the orbitals is parameterized by a unitary matrix, $\mathbf{U}^{\text{elec}}$ or $\mathbf{U}^{\text{prot}}$, which can be written as

$$|p\rangle = \sum_q |q\rangle U_{qp}^{\text{elec}} \tag{13}$$

$$|P\rangle = \sum_Q |Q\rangle U_{QP}^{\text{prot}} \tag{14}$$

where

$$\mathbf{U}^{\text{elec}} = \exp\left(\mathbf{R}^{\text{elec}}\right) = 1 + \mathbf{R}^{\text{elec}} + \frac{1}{2}\left(\mathbf{R}^{\text{elec}}\right)^2 + \ldots \tag{15}$$

$$\mathbf{U}^{\text{prot}} = \exp\left(\mathbf{R}^{\text{prot}}\right) = 1 + \mathbf{R}^{\text{prot}} + \frac{1}{2}\left(\mathbf{R}^{\text{prot}}\right)^2 + \ldots \tag{16}$$

and the elements of $\mathbf{R}^{\text{elec}}$, $R_{pk}^{\text{elec}} = -R_{kp}^{\text{elec}}$, and $\mathbf{R}^{\text{prot}}$, $R_{PK}^{\text{prot}} = -R_{KP}^{\text{prot}}$, define the independent rotation parameters. For a typical single- or multi-component CASSCF wave function, core-core, active-active, and virtual-virtual excitations are redundant and are excluded, but when a selected CI method is used to perform the CI step, the wave function is not a true CAS wave function and active-active rotations are no longer redundant.[37] Therefore, the active-active rotation parameters are included in the orbital optimization step.

The derivatives of the energy with respect to the rotation parameters define the orbital gradients, which are

$$g_{rk}^{elec} = \left(\frac{\partial E}{\partial R_{rk}^{elec}}\right)_{\mathbf{R}^{elec}=0} = 2(A_{rk} - A_{kr}) \tag{17}$$

$$g_{RU}^{prot} = \left(\frac{\partial E}{\partial R_{RU}^{prot}}\right)_{\mathbf{R}^{prot}=0} = 2(A_{RU} - A_{UR}) \tag{18}$$

with a sufficient condition for convergence of the orbitals that the orbital gradients are equal to zero.

The $A_{rk}$ are defined as

$$A_{ri} = 2F_{ri} \tag{19}$$

$$A_{ru} = \sum_v F_{rv}^{elec,core} D_{vu} + \sum_{vwx}(rv|wx)D_{vu,wx} - \sum_{vUV}(rv|UV)D_{vu,UV} \tag{20}$$

$$A_{ra} = 0 \tag{21}$$

where

$$F_{pq}^{elec,core} = h_{pq}^{elec} + \sum_i \left[2(pq|ii) - (pi|qi)\right] \tag{22}$$

$$F_{pq}^{elec} = F_{pq}^{elec,core} + \sum_{uv} D_{uv}\left[(pq|uv) - \frac{1}{2}(pu|vq)\right] - \sum_{UV} D_{UV}(pq|UV). \tag{23}$$

The $A_{RK}$ are defined as

$$A_{RU} = \sum_V F_{RV}^{prot,core} D_{VU} - \sum_{uvV}(uv|RV)D_{vu,UV} \tag{24}$$

$$A_{RA} = 0. \tag{25}$$

where

$$F_{PQ}^{prot,core} = h_{PQ}^{prot} - 2\sum_i (ii|PQ) \tag{26}$$

$$F_{PQ}^{\text{prot}} = F_{PQ}^{\text{prot,core}} - \sum_{uv} D_{uv} (uv | PQ). \tag{27}$$

The rotation parameters are updated using the Newton-Raphson step,

$$\boldsymbol{\delta}^{\text{elec}} = -\mathbf{g}^{\text{elec}} \left(\mathbf{H}^{\text{elec}}\right)^{-1} \tag{28}$$

$$\boldsymbol{\delta}^{\text{prot}} = -\mathbf{g}^{\text{prot}} \left(\mathbf{H}^{\text{prot}}\right)^{-1} \tag{29}$$

where $\boldsymbol{\delta}^{\text{elec}}$ and $\boldsymbol{\delta}^{\text{prot}}$ are vectors of rotation-parameter displacements and $\mathbf{H}^{\text{elec}}$ and $\mathbf{H}^{\text{prot}}$ are the orbital Hessians. Rather than use the exact Hessians, they are initially approximated using a diagonal guess because this matrix is trivial to invert. The elements of the electronic orbital Hessian are defined as

$$H_{rk,sl}^{\text{elec}} = (1 - \tau_{rk})(1 - \tau_{sl}) \left[ 2G_{rs}^{kl} - \delta_{kl} \left( A_{rs} + A_{sr} \right) \right] \tag{30}$$

where

$$G_{rs}^{ij} = 2 \left[ F_{rs} \delta_{ij} + L_{rs}^{ij} \right] \tag{31}$$

$$G_{rs}^{tj} = \sum_{v} D_{tv} L_{rs}^{vj} = G_{sr}^{jt} \tag{32}$$

$$G_{rs}^{tu} = F_{rs}^{c} D_{tu} + \sum_{vw} J_{rs}^{vw} D_{tu,vw} + 2 \sum_{vw} K_{rs}^{vw} D_{tv,uw} - \sum_{UV} J_{rs}^{UV} D_{tu,UV} \tag{33}$$

$$G_{rs}^{pq} = 0; \text{ otherwise,} \tag{34}$$

and

$$J_{rs}^{pq} = (rs | pq) \tag{35}$$

$$J_{pq}^{PQ} = (pq | PQ) \tag{36}$$

$$K_{rs}^{pq} = (rp | sq) \tag{37}$$

$$L_{rs}^{pq} = 4K_{rs}^{pq} - K_{sr}^{pq} - J_{rs}^{pq} \tag{38}$$

and $\tau_{rk}$ permutes orbitals $r$ and $k$.

The elements of the diagonal protonic orbital Hessian are defined as

$$H_{RK,SL}^{\text{prot}} = (1-\tau_{RK})(1-\tau_{SL})\left[2G_{RS}^{KL} - \delta_{KL}(A_{RS} + A_{SR})\right] \tag{39}$$

where

$$G_{RS}^{IJ} = 2F_{RS}\delta_{IJ} \tag{40}$$

$$G_{RS}^{TU} = F_{RS}^{c}D_{TU} - \sum_{uv} J_{uv}^{RS} D_{uv,TU} \tag{41}$$

$$G_{RS}^{PQ} = 0; \text{ otherwise} \tag{42}$$

and $\tau_{RK}$ permutes orbitals $R$ and $K$.

The electronic and protonic orbital Hessians differ slightly from the previous multicomponent CASSCF implementation[2] in that the exact diagonal orbital Hessian is used. The orbital Hessians are updated during the microiteration steps with the Broyden-Fletcher-Goldfarb-Shannon (BFGS) formula[55] using the recursive update in Ref 56. During each set of microiterations, either the electronic or protonic orbitals are optimized with the other set of orbitals held fixed.

The multicomponent HCI part of the multicomponent CASSCF method is performed using a locally modified predecessor of the Arrow program,[33-34, 50] which is designed for single-component HCI. The orbital update step is performed using a locally-modified version of the PySCF program.[57]

## 3. Results and Discussion

To assess the size of the active space required for accurate protonic densities in the multicomponent CASSCF method, a series of calculations were performed on the HCN and FHF$^-$ molecules with the hydrogen nucleus treated quantum mechanically. These two systems have emerged as the most common test systems for benchmarking new multicomponent methods. In the following, we use the notation, $(n,m)/(k,l)$, for a multicomponent CASSCF calculation with an electronic active space of $n$ electrons in $m$ orbitals and a protonic active space with $k$ protons in $l$ orbitals.

For all calculations, the cc-pVDZ and cc-pV5Z electronic basis sets[28] were used for the non-hydrogen and hydrogen atoms, respectively. The geometries of the HCN and FHF$^-$ molecules were obtained from a CCSD geometry optimization using an identical electronic basis set. We aligned the molecules on the z-axis with the classical hydrogen atom at the origin. For the quantum proton, the PB4-D protonic basis set[31] was used centered at the geometry of the classical hydrogen atom.

Using the converged multicomponent CASSCF calculations, the protonic density was computed and compared to the protonic density computed with the Fourier Grid Hamiltonian (FGH) method.[58-59] The FGH method solves the nuclear Schrödinger equation on a grid and is numerically exact for the electronically adiabatic systems in this study. The FGH calculations were performed at the single-component CCSD level of theory with identical electronic basis sets as the multicomponent CASSCF calculations. For each multicomponent CASSCF calculation, we computed the error in the protonic density as the root-mean-square error (RMSE) of the protonic density of a multicomponent CASSCF calculation relative to the FGH method as

$$\text{RMSE} = \sqrt{\sum_i^{N_{\text{grid}}} \frac{\left(\rho_i^{\text{CASSCF}} - \rho_i^{\text{FGH}}\right)^2}{N_{\text{grid}}}} \quad (43)$$

where $\rho_i$ is the density at point $i$ and the sum is over a 3D grid formed from a direct product of 32 uniformly spaced points on the x-, y-, and z-axes from 0.5806 Å to 0.6194 Å.

For the multicomponent CASSCF calculations, the initial orbitals were approximate natural orbitals obtained from a multicomponent heat-bath CISDTQ calculation using an $\varepsilon$ of $2.5*10^{-5}$ and two core orbitals. In an effort to speed up convergence, for multicomponent calculations with identical active spaces but different values of $\varepsilon$, the initial orbitals for the calculations with smaller values of $\varepsilon$ were the converged orbitals from the previous calculation with a larger value of $\varepsilon$. Additionally, for a few calculations, it was necessary to use initial orbitals from a smaller active space due to convergence issues.

Selection of the number and identity of the orbitals in the electronic and protonic active spaces was automated based on the natural orbital occupation number (NOON). We choose a series of cutoffs to determine both which electronic orbitals were treated as core orbitals and which electronic and protonic orbitals were treated as active orbitals. The cutoffs and the resulting number of core, active, and virtual orbitals for the HCN and FHF$^-$ molecules are shown in Table 1. This scheme is similar in spirit to the use of various natural orbitals in single-component quantum chemistry CASCI and CASSCF calculations.[60-64] It is possible that there exists a better algorithm for choosing an active space and further examination of the selection of the active space would be an exciting future direction for multicomponent CASSCF research. To ensure the convergence of the multicomponent CASSCF calculations with respect to $\varepsilon$, we have performed calculations with a given active space with $\varepsilon$ values of $7.5*10^{-5}$, $5.0*10^{-5}$, and $2.5*10^{-5}$.

| FHF⁻ | | HCN | |
|---|---|---|---|
| Electronic Orbitals | | Electronic Orbitals | |
| Upper NOON | Core | Upper NOON | Core |
| 2.00 | 2 | 2.00 | 2 |
| 1.99 | 4 | 1.98 | 3 |
| 1.98 | 6 | 1.97 | 4 |
| Lower NOON | Core + Active | Lower NOON | Core + Active |
| $1*10^{-2}$ | 15 | $5*10^{-3}$ | 19 |
| $5*10^{-3}$ | 21 | $1*10^{-4}$ | 27 |
| $1*10^{-4}$ | 50 | $1*10^{-5}$ | 41 |
| Protonic Orbitals | | Protonic Orbitals | |
| Lower NOON | Active | Lower NOON | Active |
| $1*10^{-3}$ | 4 | $5*10^{-4}$ | 4 |
| $1*10^{-4}$ | 6 | $1*10^{-5}$ | 7 |
| $1*10^{-5}$ | 13 | $1*10^{-6}$ | 14 |
| $1*10^{-6}$ | 21 | $1*10^{-7}$ | 22 |

**Table 1:** NOON cutoff values for the FHF⁻ and HCN molecules. For the upper NOON cutoff, a natural orbital is included in the core space if the NOON of the orbital is greater than or equal to the upper NOON cutoff value. For the lower NOON cutoff value, a natural orbital is treated as either a core or active orbital if the NOON of the orbital is greater than or equal to the lower NOON cutoff value. The number of orbitals satisfying these cutoff criteria for each molecule is listed in the second column.

In the main text, we only present a subset of the results from the multicomponent CASSCF calculations. Results from all the calculations performed in this study can be found in the Supporting Information.

We first examine the convergence of the protonic density with respect to $\varepsilon$. As can be seen in Table 2, the maximum change in the energy between any two simulations with identical active spaces and an $\varepsilon$ value of $5.0*10^{-5}$ or $2.5*10^{-5}$ is 0.8 mHa for the FHF$^-$ molecule with a (12,46)/(1,13) active space and 0.4 mHa for the HCN molecule with an (8,38)/(1,14) active space. Both of these active spaces correspond to the largest electronic active space for each molecule. The percent change in the protonic density error for these two calculations is 1.55 and 1.18 for the FHF$^-$ and HCN molecules respectively. For the FHF$^-$ molecule with a (12,27)/(1,13) active space, the difference in energy between the two calculations with an $\varepsilon$ value of $5.0*10^{-5}$ or $2.5*10^{-5}$ is 0.3 mHa with no change in the density error. For the HCN molecule with an (8,24)/(1,14) active space, the change in energy is 0.3 mHa and the percent change in the protonic density error is 0.37. The observed better convergence of the energy for a given value of $\varepsilon$ with a smaller active space is a general feature of any HCI method. Given the small changes in the protonic density error between even the calculations with the largest change in energy we conclude that the calculations presented here are sufficiently converged for the determination of the protonic density.

|         | FHF⁻      |           |               | HCN     |          |               |
|---------|-----------|-----------|---------------|---------|----------|---------------|
| $\varepsilon$ | Elec | Energy | Density Error | Elec | Energy | Density Error |
| $7.5*10^{-5}$ | (12,11) | -199.6047 | 0.695 | (8,16) | -93.0752 | 0.656 |
| $5.0*10^{-5}$ | (12,11) | -199.6047 | 0.695 | (8,16) | -93.0753 | 0.656 |
| $2.5*10^{-5}$ | (12,11) | -199.6047 | 0.695 | (8,16) | -93.0753 | 0.655 |
| $7.5*10^{-5}$ | (12,17) | -199.6794 | 0.562 | (8,24) | -93.1071 | 0.539 |
| $5.0*10^{-5}$ | (12,17) | -199.6795 | 0.561 | (8,24) | -93.1074 | 0.536 |
| $2.5*10^{-5}$ | (12,17) | -199.6796 | 0.560 | (8,24) | -93.1077 | 0.534 |
| $7.5*10^{-5}$ | (12,27) | -199.7873 | 0.330 | (8,38) | -93.1218 | 0.432 |
| $5.0*10^{-5}$ | (12,27) | -199.7876 | 0.328 | (8,38) | -93.1225 | 0.422 |
| $2.5*10^{-5}$ | (12,27) | -199.7879 | 0.328 | (8,38) | -93.1229 | 0.417 |
| $7.5*10^{-5}$ | (12,46) | -199.8080 | 0.196 | | | |
| $5.0*10^{-5}$ | (12,46) | -199.8087 | 0.193 | | | |
| $2.5*10^{-5}$ | (12,46) | -199.8095 | 0.190 | | | |

**Table 2:** Protonic density error in atomic units and absolute energies in Ha for the FHF⁻ and HCN molecules for different values of $\varepsilon$ and for different sized electronic active spaces (Elec). All FHF⁻ or HCN calculations used an active protonic space of 13 or 14 orbitals, respectively, with a single proton.

We next examine the change in the protonic density error with respect to the number of active protonic orbitals. A subset of the results is presented in Table 3. For the calculations with the two smallest electronic active spaces, we see only small changes in the protonic density error for both the FHF⁻ and HCN molecules as the number of active protonic orbitals is increased. The change is on the order of 10% for the third largest protonic active spaces, which is still not large relative to the overall errors. However, for the calculations with the largest electronic active spaces in Table 3, which correspond to 48 electronic orbitals and 39 electronic orbitals for the FHF- and HCN molecules, respectively, we see a large change in the protonic density error with respect to number of active protonic orbitals. The density error changes from 0.307 to 0.265 to 0.190 with a protonic active space of 4, 6, or 13 protonic orbitals, respectively, for the FHF$^-$ molecule and from 0.563 to 0.483 to 0.417 with a protonic active space of 4, 7, or 14 protonic orbitals, respectively, for the HCN molecule. Further increasing the size of the protonic active space for either the FHF- or HCN molecules results in only a small change in the protonic density error and a change in the energy of less than 0.1 mHa. This demonstrates that it is not necessary for a multicomponent CASSCF calculation to include all the protonic orbitals in the active space, which has not previously been shown. For a single proton, reducing the size of the protonic active space by a constant factor reduces the size of the full CI expansion by the same factor, which should help computational efficiency. For all remaining calculations, we will use a protonic active space of 13 or 14 orbitals for the FHF$^-$ or HCN molecules, respectively.

| FHF⁻ | | | | HCN | | | |
|---|---|---|---|---|---|---|---|
| Prot | Elec | Energy | Density Error | Prot | Elec | Energy | Density Error |
| 4 | (12,11) | -199.6046 | 0.698 | 4 | (8,16) | -93.0750 | 0.673 |
| 6 | (12,11) | -199.6047 | 0.695 | 7 | (8,16) | -93.0753 | 0.659 |
| 13 | (12,11) | -199.6047 | 0.695 | 14 | (8,16) | -93.0753 | 0.655 |
| 4 | (12,17) | -199.6793 | 0.572 | 4 | (8,24) | -93.1070 | 0.568 |
| 6 | (12,17) | -199.6795 | 0.564 | 7 | (8,24) | -93.1075 | 0.544 |
| 13 | (12,17) | -199.6796 | 0.560 | 14 | (8,24) | -93.1077 | 0.534 |
| 4 | (12,27) | -199.7876 | 0.349 | 4 | (8,38) | -93.1203 | 0.563 |
| 6 | (12,27) | -199.7877 | 0.343 | 7 | (8,38) | -93.1220 | 0.483 |
| 13 | (12,27) | -199.7879 | 0.328 | 14 | (8,38) | -93.1229 | 0.417 |
| 4 | (12,46) | -199.8075 | 0.307 | 22 | (8,38) | -93.1229 | 0.415 |
| 6 | (12,46) | -199.8084 | 0.265 | | | | |
| 13 | (12,46) | -199.8095 | 0.190 | | | | |
| 21 | (12,46) | -199.8095 | 0.188 | | | | |

**Table 3**: Protonic density error in atomic units and absolute energies in Ha for the FHF⁻ and HCN molecules for different sized protonic active spaces (Prot) and electronic active spaces (Elec). All calculations had a single proton in the protonic active space and used an $\varepsilon$ of $2.5*10^{-5}$.

Next, we examine the change in the protonic density error with respect to the number of core electronic orbitals. A subset of results from these calculations is presented in Table 4. As the number of core electronic orbitals controls the number of electrons in the active space, it greatly influences the number of terms in the wave function expansion in Equation 1. For the FHF⁻ molecule, the number of electronic core orbitals has a small effect on the protonic density error, with the error changing from 0.208 to 0.190 to 0.196 for 2, 4, or 6 core electronic orbitals, respectively, with an electronic combined core and active space size of 50 orbitals. Similar results are seen in the HCN calculations, where the error changes from 0.435 to 0.417 to 0.386 for 2, 3, or 4 core orbitals, respectively with an electronic combined core and active space size of 41 orbitals. From the results presented here, it appears that the number of electronic core orbitals has less of an effect on the protonic density error than the number of electronic orbitals in the combined core and electronic space.

| FHF⁻ | | | HCN | | |
|---|---|---|---|---|---|
| Elec | Energy | Density Error | Elec | Energy | Density Error |
| (16,13) | -199.6070 | 0.695 | (10,17) | -93.1645 | 0.755 |
| (16,19) | -199.7406 | 0.621 | (10,25) | -93.1645 | 0.569 |
| (16,29) | -199.8869 | 0.378 | (10,39) | -93.1857 | 0.435 |
| (16,48) | -199.9214 | 0.208 | (8,16) | -93.0753 | 0.655 |
| (12,11) | -199.6047 | 0.695 | (8,24) | -93.1077 | 0.534 |
| (12,17) | -199.6796 | 0.560 | (8,38) | -93.1229 | 0.417 |
| (12,27) | -199.7879 | 0.328 | (6,15) | -93.0179 | 0.584 |
| (12,46) | -199.8095 | 0.190 | (6,23) | -93.0374 | 0.513 |
| (8,9) | -199.5689 | 0.684 | (6,37) | -93.0470 | 0.386 |
| (8,17) | -199.5796 | 0.388 | OOMP2 | -93.1778 | 0.563 |
| (8,27) | -199.6305 | 0.307 | CCSD[31] | -93.1803 | 0.282 |
| (8,46) | -199.6425 | 0.196 | HF | -92.8452 | 1.053 |
| OOMP2 | -199.9220 | 0.271 | | | |
| CCSD[31] | -199.9254 | 0.153 | | | |
| HF | -199.4597 | 0.694 | | | |

**Table 4**: Protonic density error in atomic units and absolute energies in Ha for the FHF⁻ and HCN molecules for different sized electronic active spaces (Elec). All FHF⁻ or HCN calculations used an $\varepsilon$ of $2.5*10^{-5}$ and an active protonic space of 13 or 14 orbitals with a single proton, respectively.

In contrast to the small effect the number of electronic core orbitals has on the protonic density error, the number of active electronic orbitals has a large effect on the protonic density error. For the FHF⁻ molecule, the smallest active spaces with a given number of core orbitals have a density error essentially indistinguishable from multicomponent HF, while for the corresponding HCN calculations, the error is still very large. None of the multicomponent CASSCF calculations have a lower density error than multicomponent OOMP2 except for the calculations with the largest number of active orbitals for a given number of core orbitals. The smallest of these largest active spaces are (8,46)/(1,13) and (6,37)/(1,14) for the FHF⁻ and HCN molecules, respectively. No multicomponent CASSCF calculation in this study had a lower density error than the corresponding multicomponent CCSD calculation.

Similar to the recent studies demonstrating the accuracy of orbital-optimized multicomponent methods, this study offers further evidence of the importance of dynamic correlation for accurate protonic properties in multicomponent systems. As has been previously shown,[14] the use of multicomponent HF orbitals introduces significant multireference character into the exact multicomponent wave function, which would seem to necessitate the need for multireference methods such as multicomponent CASSCF. However, by performing the orbital-optimization procedure in the presence of electron-proton correlation, a set of electronic and protonic orbitals can be found that reduce the multireference character. Once an appropriate set of electronic and protonic orbitals has been found, dynamic correlation is essential for accurate protonic properties in a manner identical to the importance of dynamic electron-electron correlation in single-component quantum chemistry. For multicomponent methods, as long as orbital-optimization is performed in the presence of electron-proton correlation, multireference methods should not be needed except in special circumstances such as bilobal protonic densities or when static electron-electron correlation is inherent to the system in a single-component framework such as for systems containing transition metals.

## 4. Conclusions

In this study, a new two-step multicomponent CASSCF method was derived and implemented. Novel aspects of this implementation are the application of the multicomponent heat-bath CI method as the CI solver, which allows calculations with large active spaces, and the ability to perform calculations with a variable number of active protonic orbitals. The method has been benchmarked on the HCN and FHF$^-$ molecules and it was found that even with very large electronic active spaces such as (16,48) for the FHF$^-$ molecule and (10,39) for the HCN molecule, protonic densities were less accurate than with the NEO-CCSD method. Electronic active spaces larger than (8,27) or (6,15) for the FHF$^-$ and HCN molecules, respectively, are required for protonic densities that are more accurate than multicomponent OOMP2. These results confirm a long-held assumption in multicomponent quantum chemistry[24-25] that a multicomponent CASSCF calculation would need to include many electronic orbitals in the active space in order to obtain accurate protonic properties. Additionally, we find that the protonic active space can be reduced in size with a negligible change in the protonic density and the energy. This was previously unknown as prior multicomponent CASSCF studies have always included all the protonic orbitals in the active space.

The final question this study raises is how to include strong correlation in multicomponent calculations in the circumstances it is truly needed, as it would be more computationally efficient to keep the active space small, which seems difficult with a CASSCF wave function as shown in this study. Based both on the results in this and previous orbital optimized studies, we hypothesize that a restricted active space (RAS) wave function should be used with all one-electron, one-proton excitations from the reference wave function included and the strongly correlated electronic orbitals in the RAS2 subspace. Alternatively, a multicomponent multireference method with the perturbative correction included in the orbital optimization step could be used similar to the HCISCF methods of Sharma and coworkers.[35] Research towards these goals is currently underway.

## 5. Associated Content

### 5.1 Supporting Information

Energies and density errors for all combinations of $\varepsilon$ and protonic and electronic active spaces for the multicomponent CASSCF method. This information is available free of charge via the Internet at http://pubs.acs.org.

## 6. Author Information

Corresponding Author

*E-mail: brorsenk@missouri.edu

## 7. Acknowledgements

K.R.B. thanks the University of Missouri for startup funding. The computations for this work were performed on the high performance computing infrastructure provided by Research Computing Support Services and in part by the National Science Foundation under Grant No. CNS-1429294 at the University of Missouri, Columbia, MO.

**TOC Graphic**

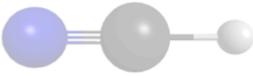
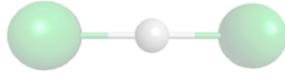

$$\left|\Psi\right\rangle = \sum_{\mu\nu} c_{\mu\nu} \left|\Phi_\mu^{\text{elec}}\right\rangle \left|\Phi_\nu^{\text{prot}}\right\rangle$$